\documentclass[a4paper,11pt,leqno,final]{amsart}

\usepackage[latin1]{inputenc}
\usepackage[T1]{fontenc}

\usepackage{amsmath}
\usepackage{showkeys}
\usepackage{amssymb}
\usepackage{amsthm}
\usepackage{color}

\usepackage[mathcal]{eucal}

\newtheorem{lemma}{Lemma}[section]
\newtheorem{prop}{Proposition}[section]

\numberwithin{equation}{section}
\newtheorem{teo}{Theorem}[section]

\renewcommand{\epsilon}{\varepsilon} 
\renewcommand{\theta}{\vartheta}
\renewcommand{\phi}{\varphi}

\newcommand{\ns}{\mathbf} 
\newcommand{\fs}{\mathcal} 
\newcommand{\cs}{\mathrm} 

\newcommand{\supp}{\mathop{\rm supp}\nolimits} 
\newcommand{\e}{\mathop{\rm e}\nolimits}

\newcommand{\di}{\, d}

\renewcommand{\(}{\left(} 
\renewcommand{\)}{\right)}

\newcommand{\vl}{\left}
\newcommand{\vr}{\right}

\newcommand{\pp}{\Big}

\renewcommand{\leq}{\leqslant} 
\renewcommand{\geq}{\geqslant}
\newcommand{\lsim}{\lesssim}
\newcommand{\gsim}{\gtrsim}

\newcommand{\prf}{\textsc{Proof.} } 
\newcommand{\bs}{~$\blacksquare$}

\newcommand{\beq}{\begin{equation}}
\newcommand{\eeq}{\end{equation}}

\newcommand{\bsp}{\begin{split}}
\newcommand{\esp}{\end{split}}

\newcommand{\bal}{ \begin{align} } 
\newcommand{\eal}{\end{align}}
\newcommand{\bals}{\begin{align*}}
\newcommand{\eals}{\end{align*}}

\newcommand{\bga}{\begin{gather}} 
\newcommand{\ega}{\end{gather}}
\newcommand{\bgas}{\begin{gather*}}
\newcommand{\egas}{\end{gather*}}

\newcommand{\dom}{ {|s-s_0| \leq t+R} } 

\begin{document}

\title[Blow up  for the semilinear Wave Equation]%
{Blow up  for the semilinear Wave Equation in Schwarzschild metric
}

\author{Davide Catania}
   \address{ Davide Catania,
Dipartimento di Matematica, Universit\`a degli Studi di Pisa,
Via F. Buonarroti 2, 56100 Pisa, Italy.}
\author{Vladimir Georgiev}
   \address{ Vladimir Georgiev,  Dipartimento di Matematica, Universit\`a degli Studi di Pisa,
Via F. Buonarroti 2, 56100 Pisa, Italy.}


\thanks{The  authors are   partially
supported by Research Training Network (RTN) HYKE,  financed by
the European Union, contract number: HPRN--CT--2002--00282. The
second author was visiting  the University of Nantes in May 2004,
where part of this work was prepared. He is grateful to prof.
Georgi Vodev for the hospitality during his visit. }

\begin{abstract}
We study the semilinear wave equation in Schwarzschild metric
($3+1$ dimensional space--time). First, we establish that the
problem is locally well--posed in $\cs H^\sigma$ for any $\sigma
\geq 1$; then we prove the blow up of the solution for every $p
\in ]1,1+\sqrt{2}[$ and non--negative non--trivial initial data.
\end{abstract}

\maketitle
\begin{centerline}
{\bf \large The work is dedicated to prof. Yvonne Choquet--Bruhat }
\end{centerline}
\begin{centerline}
{\bf \large in occasion of her 80th year. }
\end{centerline}

\section{Introduction} 

Consider the manifold
$$M = R \times \Omega,\ \mathbf
\Omega = \{ (r, \omega) : r > 2M , \ \ \omega  \in  \mathbb{S}^2
\} = (2M,\infty) \times \mathbb{S}^2,$$ equipped with the
Schwarzschild metric having the form (see chapter V in \cite{ChDD}
or chapter 31 in \cite{MTW73}):
\begin{equation}\label{eq.Schw1}
    g = F(r) \, d t^2 - F(r)^{-1} \, d r^2
  -r^2\, d \omega^2 .
\end{equation}
Here
$$
F(r) = 1 - \frac{2M}{r},$$ the constant $M>0$ has the
interpretation of mass and $\, d \omega^2$ is the standard metric
on the unit sphere $ \mathbf S^2.$

 The D'Alembert
operator associated with  the metric $g$ is
$$
\square_g = \frac{1}{F} \left( \partial^2_t - \frac{F}{r^2}
\partial_r(r^2 F) \partial_r
-\frac{F}{r^2} \Delta_{\ns S^2}\right),
$$
where $ \Delta_{\ns S^2}$ denotes the standard Laplace--Beltrami
operator on $ \ns S^2.$

Our goal is to study the existence of global solution to the
corresponding Cauchy problem for the semilinear wave
 equation
\begin{equation}\label{eq.CauchySchw1}
\square_g u = {|u|}^p  \qquad \text{in } [0,\infty[ \times \ns
\Omega.
\end{equation}

This problem can be considered as a natural analogue of the
classical semilinear wave equation
\begin{equation}\label{eq.CauchySchw1fl}
\square_{g_0} u = {|u|}^p  \qquad \text{in } [0,\infty[ \times \ns
R^n,
\end{equation}
where $ g_0 $ is the flat Minkowski metric
\begin{equation}\label{eq.M1}
    g_0 =  \di t^2 -  \di r^2
  -r^2\di \omega^2 .
\end{equation}
It is well--known (see \cite{John79}, \cite{John81}, \cite{Gla},
\cite{Sch}, \cite{Sid87}, \cite{Yi}, \cite{GLS}, \cite{JZ2} or the
review in \cite{G00} for a more complete list of references on the
subject) that for any space dimension $n \geq 2$, there exists a
critical value $p_0 = p_0(n)
>1$ such that the Cauchy problem for \eqref{eq.CauchySchw1fl} admits a global small data solution
provided $p > p_0(n).$

For subcritical values of $p \leq p_0(n)$, a blow--up phenomenon is
manifested. In the case of space dimension $n=3$, the critical
exponent is $p_0(3) = 1 + \sqrt{2},$ while in the general case of
space dimension $n \geq 2$, the critical exponent is defined as
the positive solution to
$$
(n-1) p^2 - (n+1) p -2 =0.$$ The blow up results in \cite{John79},
\cite{John81}, \cite{Gla}, \cite{Sch}, \cite{Sid87} require
a suitable comparison principle for the free wave equation. One
further remark is connected with the fact that the critical
exponent $p_0(n)$ is the same for the smaller class of radially
symmetric solutions.

The dispersive properties of the solution to the linear problem
\begin{equation}\label{eq.CauchySchw1l}
\square_g u = \Phi  \qquad \text{in } [0,\infty[ \times \ns \Omega
\end{equation}
with zero initial data, depend essentially on the distribution of
resonances for the operator
\begin{equation}\label{eq.defP}
    P =  \frac{F}{r^2}
\partial_r(r^2 F) \partial_r
+\frac{F}{r^2} \Delta_{\ns S^2}.
\end{equation}

This problem is studied in \cite{BaMo}, \cite{SaZw97} and in
\cite{BaMo}  it is shown that the resolvent $ R(z) = (z^2-P)^{-1}$
can be extended as  a meromorphic function  (as an operator from
$C_0^\infty ( \Omega)$ to $C^\infty(\Omega)$) from $\{ z \in
\mathbb{C}; \Im z > 0 \}$ to $ \mathbb{C}\setminus i\mathbb{R}.$
The result in \cite{SaZw97} shows  that the resolvent can be
extended further to a meromorphic function in the whole complex
plane  $ \mathbb{C}.$ The corresponding poles of the resolvent are
called resonances and they  are isolated and have finite rank.
Moreover, there exists a strip of type
\begin{equation}\label{eq.dom}
\{ z \in \mathbb{C}; |\Im z| < \varepsilon \}
\end{equation}
free of resonances. This phenomena is similar to the situation of
the exterior domain of several convex obstacles, studied in
\cite{Ik}, where similar domain free is resonances is found. The
approach in \cite{Ik} leads to exponential decay of the local
energy  with derivative losses. For corresponding result for
curved metrics under additional non -- trapping condition one can
see \cite{MetSogge}.

Despite of these results it seems ( according to the knowledge of
the authors) that the proof of  some concrete dispersive estimates
for the wave equation in Schwarzschild metric meets the essential
difficulty that there is no simple explicit representation of the
corresponding fundamental solution to the D'Alambert operator in
Schwarzschild metric. Similar difficulty is manifested, when one
tries to adapt the approach of F.John from \cite{John79},
\cite{John81} to the semilinear problem \eqref{eq.CauchySchw1} and
to show blow - up for some subcritical values of $p.$

Our main goal in this work is to study the semilinear wave
equation in the presence of Schwarzschild metric and to show a
blow - up result for $ 1 < p < 1 + \sqrt{2}.$

The first  step is to study the local (in time) Cauchy problem and
to show the well--posedeness in suitable Sobolev spaces.

 Introducing the Regge--Wheeler
coordinate
 \begin{equation}\label{eq.wheeler0}
    s(r) = r + 2M\log (r-2M),
\end{equation}  we can rewrite equation \eqref{eq.CauchySchw1} as (see
section \ref{sec.reform})
\begin{equation}\label{eq.eqwheeler}
   \partial^2_t u - \partial^2_s u
- \frac{2F}{r(s)}  \partial_s u -\frac{F}{{r(s)}^2} \Delta_{\ns S^2}
u = F {|u|}^p ,
\end{equation}
where $$ F=F(s) = 1 - \frac{2M}{r(s)}$$  and $r(s)$ is the function
inverse to \eqref{eq.wheeler0}.

For simplicity (and with no loss of generality), we shall restrict our considerations to the case of
solutions of the form $u = u(t,s).$ Then \eqref{eq.eqwheeler} is
simplified to the following equation:
\begin{equation}\label{eq.eqwheelersi}
   \partial^2_t u - \partial^2_s u
- \frac{2F}{r(s)}  \partial_s u = F {|u|}^p .
\end{equation}

Making further the substitution (see section \ref{sec.reform})
$$ u(t,s) = \frac{v(t,s)}{r(s)}, $$
we obtain the semilinear problem
\begin{equation}\label{eq.Hforms2}
\partial^2_t v  + G v
 = F r^{1-p} {|v|}^p ,
\end{equation}
where
\begin{equation}\label{eq.Hdefs2}
    G = - \partial^2_s  +
 \frac{2MF}{r^3}.
\end{equation}

First, we study  the local existence of the solution to the Cauchy
problem \beq \label{eq.CPnolinsi}
\begin{cases}
\partial^2_t v  + G v
 = F r^{1-p} {|v|}^p, \\
 v(0,s) = v_0(s), \qquad \partial_t v(0,s) = v_1(s).
\end{cases}
\eeq

Our next result states that the problem is locally well--posed in
$\cs H^\sigma$ for any $\sigma \geq 1.$
\begin{teo} \label{teolocalsi} Given any $\sigma \geq 1$ and any
real number $R > 0$, one can find $T=T(R)$ so that if the  initial
data
$$
v_0 \in \cs H^\sigma({\ns R}), \qquad v_1 \in \cs H^{\sigma-1}({\ns R})
$$
satisfy
$$ \|v_0\|_{\cs H^\sigma({\ns R})} + \|v_1\|_{\cs H^{\sigma-1}({\ns
R})}\leq R,$$ then the Cauchy problem \eqref{eq.CPnolinsi}  has a
unique solution
$$
v(t,s) \in \fs C^0([0,T[;\cs H^\sigma({\ns R})) \cap
\fs C^1([0,T[;\cs H^{\sigma-1}({\ns R})).$$
\end{teo}

To study the maximal time interval of existence of solutions to
the wave equation in Schwarzschild metric \begin{equation}
\label{equ}
\begin{cases}
\square_g u = |u|^p & \qquad \text{in } [0,\infty[ \times \mathbf \Omega, \\
u(0) = u_0, \quad u_t (0) = u_1 & \qquad \text{in } \mathbf
\Omega,
\end{cases}
\end{equation} we suppose that our initial data are radial
$$ u_0=u_0(r), u_1=u_1(r), \   (u_0,u_1) \in
\mathbf H^2 (\mathbf (2M,\infty)) \times \mathbf H^1 (\mathbf
(2M,\infty))
$$
and that there exists a compact interval $ \mathbf B \doteq
\overline{\mathbf B(r_0,R)} \doteq \{|r-r_0|\leq R\} \subset
\mathbf (2M,\infty),$ so that
\begin{equation} \label{data}
\begin{cases}
u_0 (r), \ u_1 (r) \geq 0 & \qquad \text{almost everywhere}, \\
    u_0 (r) = u_1 (r) = 0 & \qquad \text{for }   \ |r-r_0|\geq R, \\
\int_{2M}^\infty u_j (r) \, d r \geq \epsilon & \qquad j=0,1
\end{cases}
\end{equation}
for  positive constants $\epsilon$, $R>0$ and $r_0=r_0(\epsilon,p)
\in \Omega$. We also assume that $r_0$ is near $2M$ for $p \in
]2,1+\sqrt{2}[$, far from it for $p \in ]1,2[$ (we make no
assumption in the case $p=2$).

Now we can state the main result.

\begin{teo}\label{teou}  For any $p,\  1 < p <
1+\sqrt{2}$ there exists a positive number $\varepsilon_0$ so that
for any $\varepsilon \in (0, \varepsilon_0)$ there exists
$r_0=r_0(p,\varepsilon)$ and $R=R(p,\varepsilon) $ so that for any
initial data
$$ u_0=u_0(r), u_1=u_1(r), \   (u_0,u_1) \in
\mathbf H^2 (\mathbf (2M,\infty)) \times \mathbf H^1 (\mathbf
(2M,\infty))
$$
satisfying (\ref{data}) in $ B \doteq \overline{\mathbf
B(r_0,R)}$, there exists a positive number
$T=T(\varepsilon)<\infty$ and a solution
$$u \in \cap_{k=0}^2 \mathcal C^k ([0,T[;\mathbf H^{2-k} ((2M,\infty))$$ of
 (\ref{equ}) such that
 $$
\lim_{t \nearrow T} ||u(t)||_{\mathbf L^2((2M,\infty))} = \infty.
$$
\end{teo}

The above result means that the wave equation in Schwarzschild
metric has a similar critical exponent as the free wave equation.
However, the proof we follow here suggests  that the lifespan of
the solution has a completely different behavior
--- it might be  much longer than in the corresponding flat case ( see lemma \ref{l.Kn} below).

The main difficulty to establish the blow up of the solution is
connected with the sign changing  properties of  the fundamental
solution of the linear wave equation in Schwarzschild metric (or
more generally   in curved metrics). For the case of flat $1+3$
Minkowski metric the fundamental solution is non - negative and
this property is used effectively in the study of the blow - up
phenomena for the corresponding semilinear wave equation.

The proof of the main theorem is based on the application of a variant of the classical Kato's
lemma (see lemma \ref{l.K} below) for an average of type
\begin{gather}  \label{defF1n}
\int_{\ns R}  v(t,s) \psi_0 (s) \di s,
\end{gather}
where  $v$ is a solution to \eqref{eq.Hforms2} and $\psi_0 =
\phi_0 - D > 0$, with $\phi_0$ being a suitable  solution to
\begin{alignat}{2}\label{eq.eleq}
&\left(- \partial^2_s  +
 \frac{2MF}{r^3}\right)\phi_0 = 0.
\end{alignat}
In the flat case (i.e. in the case $M=0$), it is trivial that
\eqref{eq.eleq} has a bounded solution (all constants are bounded
solutions). However, in the case of $M > 0$, we shall find a
special solution satisfying the following conditions at infinity:
\beq \vl\{
\begin{array}{ll} \label{eq.asimp}
   |\phi_0 (s) - bs| \lsim \log(2+s) & \quad \text{for $s\geq 0$}, \\
   \phi_0 (s) \sim |s|\e^s/2M  & \quad \text{if } s \rightarrow -\infty
\end{array} \vr.
\eeq (see section \ref{mainproof} for the notations). This
different asymptotic behavior of the solution to the elliptic
equation $G \varphi_0 =0$ interacts with the factor $F$ in the
right side of equation \eqref{eq.Hforms2}.

In this way, we are able to handle the domain near the black hole
and to show that a suitable modification of the classical
Kato's lemma works well.

\section{Reformulation of the problem and local existence result.} 
\label{sec.reform}

The D'Alembert operator in the metric \eqref{eq.Schw1} has the
form
\begin{equation}\label{eq.dalemb}
    \square_g  = \frac{1}{F} \partial^2_t - F \partial^2_r
- \frac{2}{r} \left( 1-\frac{M}{r} \right) \partial_r
-\frac{1}{r^2} \Delta_{\ns S^2} , \ \ F = 1 - \frac{2M}{r}.
\end{equation}

Introducing the Regge--Wheeler  coordinate
\begin{equation}\label{eq.wheeler}
    s(r) = r + 2M\log (r-2M),
\end{equation}
we have the relation
$$
\frac{r}{r-2M} \di r = \di s
$$
and the metric \eqref{eq.Schw1} becomes
\begin{equation}\label{eq.Schw1we}
    g = F \( \di t^2 - \di s^2
  -\Lambda(s)^2\di \omega^2 \),
\end{equation}
where
\begin{equation}\label{eq.comffac}
  \Lambda(s) = \frac{r(s)}{\sqrt{F(s)}} = \frac{r(s)}{\sqrt{1-\frac{2M}{r(s)}}}
\end{equation}
and $r(s)$ denotes the function inverse to \eqref{eq.wheeler}.
Hence the Schwarzschild metric $g$ is conformal to the metric $
\di t^2 - \tilde{g},$ where $\tilde{g}$ is the following metric
\begin{equation}\label{eq.Schw1wen}
    \tilde{g} =  \di s^2
  + \Lambda(s)^2\di \omega^2
\end{equation}
on  the cylinder $\ns M = {\ns R} \times {\ns S^2}.$

The Laplace--Beltrami operator associated with the metric $\tilde{g}$
has the form
\begin{equation}\label{eq.hlb}
    \Delta_{\tilde{g}} =  \partial_s^2  + \frac{2}{r} \left( 1 - \frac{3M}{r}
    \right)\partial_s + \frac{1}{\Lambda^2(s)}\Delta_{\ns S^2}.
\end{equation}

In the sequel, we shall need the following asymptotic estimates
for the functions $r(s)$, $F(s)$, $\Lambda (s)$ (which follow
trivially from \eqref{eq.wheeler}).

\begin{lemma}\label{lm,asrfl}
There exist positive constants $C_2 >C_1 >0$ such that the
functions $$r(s), \qquad F(s) = 1 - \frac{2M}{r(s)}, \qquad
  \Lambda(s) = \frac{r(s)}{\sqrt{F(s)}}$$ satisfy the estimates
\begin{equation}\label{eq.estrs}
\left\{%
\begin{array}{ll}
   C_1  s \leq r(s) \leq C_2  s & \quad \hbox{if \ $s \geq 2$,} \\
   C_1  \leq  r(s) \leq C_2 & \quad \hbox{if \ $|s| \leq 2$,} \\
   C_1 \e^{s/2M} \leq r(s)-2M \leq C_2 \e^{s/2M} & \quad \hbox{if \  $s \leq -2$;} \\
\end{array}%
\right.
\end{equation}
\begin{equation}\label{eq.estFs}
\left\{%
\begin{array}{ll}
    |F(s)-1| \leq C_2  /s & \quad \hbox{if \ $s \geq 2$,} \\
   C_1  \leq  F(s) \leq C_2 & \quad \hbox{if \ $|s| \leq 2$,} \\
   C_1 \e^{s/2M} \leq F(s) \leq C_2 \e^{s/2M} & \quad \hbox{if\  \ $s \leq -2$;} \\
\end{array}%
\right.
\end{equation}
\begin{equation}\label{eq.estlas}
\left\{%
\begin{array}{ll}
   C_1 s \leq \Lambda(s) \leq C_2  s & \quad \hbox{if \ $s \geq 2$,} \\
   C_1  \leq  \Lambda(s) \leq C_2 & \quad \hbox{if \ $|s| \leq 2$,} \\
   C_1 \e^{-s/4M} \leq \Lambda(s) \leq C_2 \e^{-s/4M} &
       \quad \hbox{if \  $s \leq -2$.} \\
\end{array}%
\right.
\end{equation}

\end{lemma}

Further, we have the relations $F\partial_r = \partial_s$ and
\begin{align*}
\partial_s^2 & = F \partial_r (F \partial_r) = \( 1 - \frac{2M}{r} \) \partial_r
                    \( \(1 - \frac{2M}{r}\) \partial_r \) \\
             & = \frac{2M}{r^2} \( 1 - \frac{2M}{r} \) \partial_r
                    + {\( 1 - \frac{2M}{r} \)}^2 \partial_r^2,
\end{align*}
and this implies \beq F^2 \partial_r^2 =
\partial_s^2 -\frac{2M}{{r(s)}^2} \partial_s.
\eeq

In this way, from \eqref{eq.dalemb} we obtain the relation
\begin{equation}\label{eq.dalembs}
   F \square_g  = \partial^2_t - \partial^2_s
- \frac{2F}{r(s)}  \partial_s -\frac{F}{r(s)^2} \Delta_{\ns S^2}
 = \partial_t^2 -   \Delta_{\tilde{g}} - \frac{2M}{r^2} \partial_s
\end{equation}
and the problem \eqref{eq.CauchySchw1} takes the form
\eqref{eq.eqwheeler}, i.e.
\begin{equation}\label{eq.eqwheelerag}
   \partial^2_t u - \partial^2_s u
- \frac{2F}{r(s)}  \partial_s u -\frac{F}{r(s)^2} \Delta_{\ns S^2}
u = F |u|^p.
\end{equation}

Now, we make the change
\begin{equation}\label{eq.conff}
    u = w v,
\end{equation}
where
$$
w = w(s)$$ will be chosen appropriately. We have the relations
$$
\partial_s (w v) = v \partial_s w + w \partial_s v, $$
$$
\partial^2_s (w v) = v \partial^2_s w + 2 \partial_s v \partial_s w + w \partial^2_s v ,$$
and from \eqref{eq.eqwheelerag} we obtain the equation \beq
\label{eq.wv1}
\begin{split}
w \partial^2_t v &- w\partial^2_s v
    - 2\left( \partial_s w + \frac{F}{r(s)}w \right)  \partial_s v \\
  & - \left( \partial^2_s w + \frac{2F}{r(s)} \partial_sw \right) v
    - \frac{F}{r(s)^2} w \Delta_{\ns S^2} v \\
  &= F|w|^p |v|^p .
\end{split}
\eeq

If we want to cancel the coefficient $$\partial_s w +
\frac{F}{r(s)}w, $$ then a simple computation shows that we have
to take
\begin{equation}\label{eq.chw}
   w(s) = \frac{1}{r(s)};
\end{equation}
moreover, with this choice, we have
$$
\partial^2_s w = \frac{2F}{r(s)^2} \left( 1 -
\frac{3M}{r(s)}\right) w.$$ These calculations enable us to
rewrite \eqref{eq.wv1} as follows:
\begin{eqnarray}
   \partial^2_t v - \partial^2_s v +
 \frac{2MF}{r^3} v
- \frac{F}{r^2}  \Delta_{\ns S^2} v = F|w|^{p-1} |v|^p .
\label{eq.wv1v}
\end{eqnarray}

At this point, with no loss of generality, we can restrict our
attention to the class of solutions of the form $v = v(t,s)$. Then
equation \eqref{eq.wv1v} can be simplified further to the
following one:
\begin{equation}\label{eq.Hform}
\partial^2_t v  + G v
 = F r^{1-p} |v|^p ,
\end{equation}
where
\begin{equation}\label{eq.Hdef}
    G = - \partial^2_s  +
 \frac{2MF}{r^3};
\end{equation}
therefore, if we set $f = f(s) = F(s){r(s)}^{1-p}$, theorem
\ref{teou} is equivalent to the following.

\begin{teo} \label{teov}
Let consider the Cauchy problem \beq \label{probv}
\begin{cases}
v_{tt} + Gv = f{|v|}^p & \qquad \text{in } [0,T[ \times \ns R, \\
v(0,s) = v_0(s), \quad v_t(0,s) = v_1(s) & \qquad \text{in } \ns
R,
\end{cases}
\eeq with initial data $(v_0,v_1) \in \cs H^2 (\ns R) \times \cs
H^1 (\ns R)$ such that \beq \label{ipdi}
\begin{cases}
v_0 (s), \ v_1 (s) \geq 0 & \qquad \text{almost everywhere}, \\
    v_0 (s) = v_1 (s) = 0 & \qquad \text{for } |s - s_0| > R, \\
\int_{\ns R} v_j (s) \di s \geq \epsilon & \qquad j=0,1
\end{cases}
\eeq for a positive constant $R$ and $s_0 = s_0 (\epsilon,p) \in \ns R$ so that, for $p \not = 2$, we have
$$
\lim_{\epsilon \rightarrow 0} \epsilon^A |s_0(\epsilon,p)| = 0 \qquad \forall A > 0
$$
and
$$
\begin{cases}
s_0 >0 & \qquad \text{for } p \in ]1,2[, \\
s_0 <0 & \qquad \text{for } p \in ]2, 1+\sqrt{2}[.
\end{cases}
$$

Then, for each $p \in ]1,1+\sqrt{2}[$, there exists a positive number $T < \infty$
and a solution$$v \in \fs   \cap_{k=0}^2 \fs C^k ([0,T[;\cs
H^{2-k} (\ns R))$$ of \eqref{probv} such that
$$
\lim_{t \nearrow  T} \vl(  \| v(t)\|_{\cs H^2(\ns R)} +
\|\partial_t v(t)\|_{\cs H^1(\ns R)}
 + \|\partial^2_t v(t)\|_{\cs L^2(\ns R)} \vr) = \infty.
$$
\end{teo}

More precisely, we shall show that
$$
\lim_{t \nearrow  T} \| v(t)\|_{\cs L^2(\ns R)} = \infty
$$
(see section \ref{mainproof}).

It is not difficult to see that $G$ is a non--negative  symmetric
operator in the Hilbert space $\cs L^2( {\ns R}, ds ) $ with dense
domain $\cs H^2( {\ns R}).$ Thus the estimates of lemma
\ref{lm,asrfl}, together with the KLMN--theorem (see theorem $
10.17 $ in \cite{ReSi}), imply that $G$ is a non--negative
self--adjoint operator.

Indeed, one can
 consider the quadratic form
\begin{eqnarray}
        B(h,h) = (\partial_s h,\partial_s h)_{\cs L^2({\ns R}, ds)}
               + \int_{{\ns R}} \frac{2MF}{r^3}{|h(s)|}^2 \di s.
    \end{eqnarray}
To apply the KLMN--theorem,  it is sufficient to verify the
estimate
\begin{eqnarray} \label{eq.klmn}
        \int_{{\ns R}} V(s) |h(s)|^2 \di s \leq
        a \int_{{\ns R}} |\partial_s h(s)|^2 \di s
          + b\|h\|^2_{\cs L^2({\ns R}, ds)}
 \end{eqnarray}
with $0 \le a < 1$, $$  V(s) = \frac{2MF(s)}{r(s)^3} $$ and $ h
\in \cs H^1({\ns R})  $. This estimate follows from lemma
\ref{lm,asrfl} with $a=0$.

Let consider the Cauchy problem \beq\label{eq.CPlin}
\begin{cases}
\partial^2_t v  + G v
 = \Phi, \\
 v(0,s) = v_0(s), \quad \partial_t v(0,s) = v_1(s);
\end{cases}
\eeq then the solution can be represented in the form
$$
v(t) = \cos(t\sqrt{G}) v_0 + \frac{\sin(t\sqrt{G})}{\sqrt{G}}v_1 +
\int_0^t \frac{\sin((t-\tau)\sqrt{G})}{\sqrt{G}}\Phi(\tau) d\tau.
$$

From this representation, we find \beq\label{eq.est0}
\begin{split}
    \|v(t)\|_{\cs L^2({\ns R}, ds)} &\leq \|v_0\|_{\cs L^2({\ns R}, ds)}
      + t\|v_1\|_{\cs L^2({\ns R}, ds)} \\
         &\qquad + \int_0^t |t-\tau|\|\Phi(\tau)\|_{\cs L^2({\ns R}, ds)} \di \tau
\end{split}
\eeq and for any $\sigma \ge 1$ we have \beq\label{eq.estsi}
\begin{split}
    \|G^{\sigma/2}v(t)\|_{\cs L^2({\ns R}, ds)} &\leq
        \|G^{\sigma/2}v_0\|_{\cs L^2({\ns R},ds)}
        + \|G^{(\sigma-1)/2}v_1\|_{\cs L^2({\ns R}, ds)} \\
    & \qquad +\int_0^t \|G^{(\sigma-1)/2}\Phi(\tau)\|_{\cs L^2({\ns R}, ds)}
    \di \tau;
\end{split}
\eeq thus, using the equivalence
$$
\|G^{\sigma/2}h\|_{\cs L^2({\ns R}, ds)} + \|h\|_{\cs L^2({\ns R},
ds)} \sim \|h\|_{\cs H^\sigma({\ns R})},
$$
we  arrive at the energy estimate \beq
\begin{split}
    \|v(t)\|_{\cs H^\sigma({\ns R})} &\leq C\|v_0\|_{\cs H^\sigma({\ns R})}
             + C(1+t)\|v_1\|_{\cs H^{\sigma-1}({\ns R})} \\
    & \qquad +C\int_0^t (1+|t-\tau|)\|\Phi(\tau)\|_{\cs H^{\sigma-1}({\ns R})}
                   \di \tau
\end{split}
\eeq for any $\sigma \geq 1$ and a suitable constant $C>0$. From
this energy estimate and the Sobolev embedding $\cs H^1({\ns R})
\subset \cs L^\infty({\ns R})$, we easily obtain the desired local
existence result (theorem \ref{teolocalsi}).

\section{Blow up with small initial data for $p \in ]2,1+\sqrt{2}[$.} 
\label{mainproof}

We shall suppose
\begin{equation}\label{eq.p12}
    2 < p < 1 +\sqrt{2}
\end{equation}
and consider the Cauchy problem \beq \label{probva}
\begin{cases}
v_{tt} + Gv = f{|v|}^p & \qquad \text{in } [0,T[ \times \ns R, \\
v(0,s) = v_0(s), \quad v_t(0,s) = v_1(s) & \qquad \text{in } \ns
R,
\end{cases}
\eeq
 with initial data $(v_0,v_1) \in \cs H^2 (\ns R) \times \cs
H^1 (\ns R)$ such that \beq \label{ipdi12}
\begin{cases}
v_0 (s), \ v_1 (s) \geq 0 & \qquad \text{almost everywhere}, \\
    v_0 (s) = v_1 (s) = 0 & \qquad \text{for } |s - s_0| > R, \\
\int_{\ns R} v_j (s) \di s \geq \varepsilon & \qquad j=0,1
\end{cases}
\eeq for a positive constant $\varepsilon$, $R>0$ and $s_0 \in \ns
R$. We shall choose suitably $s_0 = s_0(\varepsilon)<0$ in
accordance with lemma \ref{l.Kn}.

\begin{teo} \label{teovda}
If $p \in ]2,1+\sqrt{2}[$ and the initial data $(v_0,v_1) \in \cs
H^2 (\ns R) \times \cs H^1 (\ns R)$ satisfy the assumptions
\eqref{ipdi12} with
\begin{equation}\label{eq.sep}
    s_0 = s_0(\varepsilon) < 0
\end{equation}
such that
\begin{equation}\label{eq.sc0}
 \lim_{\varepsilon \searrow 0} \varepsilon^A  |s_0(\varepsilon)|
 = \infty \qquad \forall A >0,
\end{equation}
 then there exists a positive number $T < \infty$ and a
solution
$$v \in \fs \cap_{k=0}^2 \fs C^k ([0,T[;\cs H^{2-k} (\ns R))$$ of
\eqref{probv} such that
$$
\lim_{t \nearrow  T} \vl(  \| v(t)\|_{\cs H^2(\ns R)} +
\|\partial_t v(t)\|_{\cs H^1(\ns R)}
 + \|\partial^2_t v(t)\|_{\cs L^2(\ns R)} \vr) = \infty.
$$
\end{teo}

As we shall prove in the next section (lemmas \ref{lemfi0} and
\ref{lemfi1}), there exist two positive functions $\phi_0$ and
$\phi_1$ belonging to $\fs C^2 (\ns R)$ with the following
properties:
\begin{gather*}
G\phi_0 = 0, \qquad \begin{cases}
\phi_0(s) \sim s & \quad \text{for $s\rightarrow \infty$}, \\
 \phi_0(s)-D \sim \e^{s/2M} & \quad \text{for }
 s \rightarrow -\infty;
\end{cases} \\
(G+1/4M^2)\phi_1 = 0, \qquad \phi_1 (s) \sim \e^{s/2M} \quad
\text{if } |s| \rightarrow \infty,
\end{gather*}
where $D$ is a positive constant.

Here and below, we shall use the notation
$$ f(s) \sim g(s) \qquad \text{if } s \rightarrow \pm \infty,$$
where $f(s)$, $g(s)$ are given  functions, if there exist two
positive constants $C >1$ and $N >0$ such that $g(s) \neq 0$ for
$\pm s \geq N$ and
$$
C^{-1} < \frac{f(s) }{g(s)} < C  $$ for $\pm s \geq N.$

Similarly,
$$ f(s) \sim g(s) \qquad \text{if } |s| \rightarrow  \infty$$
means
$$ f(s) \sim g(s) \qquad \text{if } s \rightarrow + \infty$$
and
$$ f(s) \sim g(s) \qquad \text{if } s \rightarrow - \infty.$$

On the other hand, the notation
$$ f(s) \lesssim g(s),$$
where $f(s)$, $g(s)$ are given  functions, means that  there
exists a positive constant $C$ such that
$$
 f(s) \leq C g(s)  $$ for all $ s.$

Let $$v \in \fs   \cap_{k=0}^2 \fs C^k ([0,T[;\cs H^{2-k} (\ns R))$$ be a solution
of (\ref{probv}); we set (see \cite{JZ1}, \cite{JZ2})
\begin{gather}  \label{defF1}
F_0 (t) = \int_{\ns R} v(t,s) \psi_0 (s) \di s, \quad F_1 (t) =
\e^{-t/2M} \int_{\ns R}  v(t,s) \phi_1 (s) \di s,
\end{gather}
where
\begin{equation}\label{eq.psi0}
   \psi_0 = \varphi_0 - D > 0.
\end{equation}

We have the following relations:
\begin{equation}\label{eq,psi0a}
   \partial_s^2 \psi_0- \frac{2MF}{r^3}\psi_0 = D \frac{2MF}{r^3},
\end{equation}
\begin{equation}\label{eq.psi0b}
    \psi_0(s) \sim \begin{cases}
     s  & \quad \text{for } s \rightarrow \infty, \\
     \e^{s/2M} & \quad \text{for } s \rightarrow -\infty.
    \end{cases}
\end{equation}

Note that  the above functions $F_0, F_1$ are well--defined
(thanks to the assumptions on the initial data) and $F_0 \in \fs C^2$
due to the assumption
$$v \in \fs   \cap_{k=0}^2 C^k ([0,T[;\cs H^{2-k} (\ns R)).$$
Applying the technical  lemma \ref{lemF1} from the appendix (section 4), we find
$F_1 (t) \gsim \epsilon$.

We shall use in the sequel  the following classical Kato's lemma
(see \cite{K80}).

\begin{lemma}\label{l.K}
Suppose that $p>1$, $a \geq 1,$  and $(p-1)a > q-2$. If $V \in \fs
C^2 ([0,T[)$ satisfies
\begin{align}
& V (t) \gsim {(t+R)}^a , \label{estF0n} \\
& V'' (t) \gsim {(t+R)}^{-q} {V(t)}^p \label{estF0''n}
\end{align}
for a positive constant $R$, then $T < \infty$.
\end{lemma}

We shall need the following variant of this lemma.

\begin{lemma}\label{l.Kn}
Suppose that $p>1$, $a \geq  1 ,$ $\varepsilon \in ]0,1[$  and
\begin{equation}\label{eq.paq}
   (p-1)a > q-2.
\end{equation}
 We assume that $V(t) \in \fs C^2 ([0,T[)$ and $U(t) \in \fs C ([0,T[)$ are
non--negative functions that satisfy the following estimates
\begin{align}
& U(t)^p \gsim {(t+R)}^{-q} {V(t)}^p + \varepsilon^p (t+R)^{a-2}, \label{est-1} \\
& V (0) \gsim \varepsilon ,\qquad V^\prime (0) \gsim \varepsilon, \label{estF0ep} \\
& V'' (t) \gsim {U(t)}^p - C U(t)\label{estF0''}
\end{align}
for $t \in ]0,T[.$ Suppose further that there exists  a positive
number  $\delta \in ]0,1[$  and there exists
$T_1=T_1(\varepsilon) < T$ such that for any positive $A>0$ we
have
\begin{equation}\label{eq.lim}
    \lim_{\varepsilon \rightarrow 0} \varepsilon^A T_1(\varepsilon) = \infty
\end{equation}
and
\begin{align}
& V'' (t) \gsim  {U(t)}^p - \e^{-\delta T_1} U(t) \qquad
\text{for } t \in [0,T_1[ \label{estF0''t1};
\end{align}
 then, for any integer $N \geq 1$, we have
\begin{equation}\label{estF0}
    V(t) \gtrsim  (t+R)^N \qquad \text{for } t \in
    ]T_1/2,T_1[,
\end{equation}
and $T < \infty$.
\end{lemma}

\prf Assume that $T = \infty.$ Our goal is to arrive at a contradiction.
Consider the function
\begin{equation}\label{eq.spomf}
   K(x) = x^p - C x,
\end{equation}
where $C>0$ is the constant from \eqref{estF0''}. For
\begin{equation}\label{eqx0}
    x > x_0 \doteq 2 C^{1/(p-1)},
\end{equation}
 we have
\begin{equation}\label{eq.ozotd}
    K(x) \gtrsim x^p.
\end{equation}
In a similar way, given $T_1>0$, we can consider the function
\begin{equation}\label{eq.spomft1}
   K_{T_1}(x) = x^p - \e^{-\delta T_1} x,
\end{equation}
where  $\delta>0$ is the constant from \eqref{estF0''t1}. For
\begin{equation}\label{eq.inla}
   x
> x_0(T_1)  \doteq 2 \e^{-\delta T_1/(p-1)},
\end{equation}
 we have
\begin{equation}\label{eq.ozotdt1}
    K_{T_1}(x) \gtrsim x^p.
\end{equation}

Note that inequality \eqref{est-1} assures that
\begin{equation}\label{eq.ozot1}
    U(t) \gsim  \varepsilon (t+R)^{(a-2)/p} \gsim \varepsilon
    (T_1+R)^{(a-2)/p}
\end{equation}
for  $t \in [0,T_1]$
if $a \leq 2$, and
\begin{equation}\label{eq.ozot2}
    U(t) \gsim  \varepsilon (t+R)^{(a-2)/p} \gsim \varepsilon
    R^{(a-2)/p}
\end{equation}
for  $t \in [0,T_1]$
if $ a > 2.$ Now it is clear that choosing
$T_1=T_1(\varepsilon)>0$ so that \eqref{eq.lim} is satisfied, we
can guarantee the following analogue of inequality
\eqref{eq.inla}:
\begin{equation}\label{eq.inlat}
   U(t)
> x_0(T_1)  \doteq 2 \e^{-\delta T_1/(p-1)} \qquad  \text{for } t \in
[0,T_1].
\end{equation}

Indeed, the lower bound of $U(t)$ is at most polynomially decaying
(in $T_1$)  due to \eqref{eq.ozot1} and \eqref{eq.ozot2}, while $
x_0(T_1)$ decays exponentially.

Since \eqref{eq.inla} implies \eqref{eq.ozotdt1}, we conclude that
$$
 V'' (t) \gsim  {U(t)}^p  \geq \varepsilon^p (t+R)^{a-2} \qquad
\text{for } t \in [0,T_1[, $$ and integrating this inequality
twice, we obtain
\beq \label{est.VV'T1}
V(t) \gsim \varepsilon^p (t+R)^{a}, \quad V^\prime(t) \gsim \varepsilon^p (t+R)^{a-1} \quad \text{for } t \in ]T_1/2,T_1[.
\eeq Now we apply
\eqref{est-1} and find
\begin{equation}\label{eq.ut}
    U(t)^p \gtrsim (t+R)^{ap-q} \varepsilon^{p^2} .
\end{equation}
Setting
\begin{equation}\label{eqr1}
    a_1 = ap - q +2, \qquad p_1 = p^2,
\end{equation}
we get the following analogue of \eqref{est-1}:

$$ U(t)^p \gsim {(t+R)}^{-q} {V(t)}^p + \varepsilon^{p_1}
(t+R)^{a_1-2}.$$ Note that the assumption $a(p-1) > q-2$ implies
\begin{equation}\label{eq.10}
a_1 >a.
\end{equation}

Repeating this argument, we define the following recurrence
sequence:
\begin{equation}\label{eq.recc}
   a_0=a, \qquad  a_{k+1} = p a_k - q +2, \qquad p_{k+1} = p_k^2
\end{equation}
for any $k \geq 0$, and obtain
\begin{equation}\label{eq.utka}
\begin{cases}
U(t)^p \gsim {(t+R)}^{-q} {V(t)}^p + \varepsilon^{p_{k+1}}
(t+R)^{a_{k+1}-2}, \\
V(t) \gsim \varepsilon^p (t+R)^{a_{k+1}}.
\end{cases}
\end{equation}

It is not difficult to see that $a_k$ tends to infinity. In fact,
\eqref{eq.recc} implies
\begin{equation}\label{eq.gpr}
    a_{k+1} - a_k = p ( a_k - a_{k-1}) \geq \cdots \geq p^k (a_1 -
    a_0);
\end{equation}
so, from \eqref{eq.10} and the assumption $p>1$, we see that
$$
a_k \longrightarrow  \infty. $$ This proves \eqref{estF0}.
 We can choose in particular $k
\geq 1$ so that $a_{k+1} > 2 $ and fix this $k.$ Then estimate
\eqref{eq.utka} implies
\begin{equation}\label{eq.lele}
    U(T_1) > 2 x_0,
\end{equation}
where $x_0$ is the constant chosen in \eqref{eqx0}. Once this
inequality is satisfied, we can use \eqref{est-1}, \eqref{estF0''}
and \eqref{eq.lele},  and see that
\begin{align}
& U(t)^p \gsim {(t+R)}^{-q} {V(t)}^p + \varepsilon^p (t+R)^{a-2}\label{est-1n}, \\
& U(T_1) > 2 x_0, \quad V(T_1) > 2 x_0, \quad  V^\prime(T_1) \geq 0, \label{estF0epn} \\
& V'' (t) \gsim {U(t)}^p - C U(t)\label{estF0''nn}
\end{align}
for $t \in [T_1, 2T_1].$ Let
$$
T_2 = \sup \{T > T_1\; ; \; U(t) > x_0 , \; t \in [T_1,T[\}.$$
One can see that $T_2 \geq 2T_1.$ Indeed, for $t \in [T_1,T_2[$, we
have
$$
V'' (t) \gsim {U(t)}^p;$$ hence, using inequality \eqref{est-1n}
 and integrating twice in $t$, we conclude that \eqref{eq.utka}
 are fulfilled for $t \in [T_1,T_2[$. Thus, at $t = T_2$, we have
$$
U(T_2) > 2x_0$$ and this inequality, combined with the definition
of $T_2$, guarantees that $T_2 \geq 2T_1.$

Since \eqref{eq.utka} are fulfilled for $t \in [T_1,2T_1]$, we are
ready to obtain the desired contradiction. To see this, it is
sufficient to note that the ordinary differential equation
$$
v'' (t)  \geq v^r,$$  with initial data
$$ v(T_1) \geq 1, \qquad v^\prime(T_1) \geq 0$$ and $1 < r < p$, has a finite lifespan.

  This completes the proof.
 \bs

\vspace{.5cm}

We want to apply this lemma with $V=F_0$,
$$U =  \left( \int  f{|v|}^p \psi_0 \di s\right)^{1/p},$$
$p$ and $R$ as stated in theorem \ref{teovda}, $q=3(p-1)$ and with
$a=4-p-\epsilon_0$ for each $\epsilon_0>0$. Note that
this choice guarantees that the inequality
$$
(p-1)a -q + 2 >0$$ is equivalent to
$$
p^2 - (2-\epsilon_0)p - (1+\epsilon_0) >0$$ and this, for $p > 1$,
means that
$$
p < 1 + \sqrt{2+\frac{\epsilon_0^2}{4}} -\frac{\epsilon_0}{2};
$$
hence, choosing $\epsilon_0$ arbitrarily small, we get our result
for every $$p \in ]2,1+\sqrt{2}[.$$

First of all, we observe that the conditions in \eqref{estF0ep} are trivially satisfied. Then, in order to prove estimate \eqref{estF0''}, we multiply equation \eqref{probv}
by $\psi_0$, integrate on $\ns R$ and then integrate by parts:
\begin{gather*}
\begin{split}
F_0'' (t) &=    \int_{\ns R}  v_{tt} (t,s) \psi_0 (s) \di s \\
            &=    \int  \( v_{ss} -\frac{2MF}{{r(s)}^3} v \)
                       \psi_0 \di s
                  + \int  f{|v|}^p \psi_0 \di s\\
            &=    \int  v \( \psi_0'' - \frac{2MF}{{r(s)}^3} \psi_0\) \di s
                  + \int  f{|v|}^p \psi_0 \di s; \\
\end{split}
\end{gather*}
from \eqref{eq,psi0a}, we obtain
 \begin{align} \begin{split} \label{F0''nonlin} F_0'' (t) & = \int f{|v|}^p
 \psi_0 \di s + D\int  v  W(s) \di s  \\
 & = {U(t)}^p + D\int  v W(s)  \di s, \end{split}
 \end{align}
where
$$
W(s) = \frac{2MF}{{r(s)}^3}.
$$

 Now let note that as $r$ goes from $1$ to $\infty$, $s$ varies in the whole real line.
 Moreover, $s(r)$ is strictly increasing, thus the same holds for $r = r(s)$ and from
 \eqref{eq.estrs} we have
\begin{align} \label{relrs}
r = r(s) \sim \vl\{
\begin{array}{ll}
s & \quad \text{if } s \rightarrow \infty, \\
2M + \e^{s/(2M)} & \quad \text{if } s \rightarrow -\infty,
\end{array} \vr.
\end{align}
which yields

\beq \label{finfty} f(s) = \frac{F}{{r(s)}^{p-1}} \sim
\begin{cases}
s^{-(p-1)} & \quad \text{if } s \rightarrow \infty, \\
\e^{s/(2M)} & \quad \text{if } s \rightarrow -\infty,
\end{cases}
\eeq

\begin{equation}\label{eq.ws}
W(s) = \frac{2MF}{{r(s)}^3} \sim
\begin{cases}
s^{-3} & \quad \text{if } s \rightarrow \infty, \\
\e^{s/(2M)} & \quad \text{if } s \rightarrow -\infty
\end{cases}
\end{equation}
and in particular
 \beq \label{est.fpsiW}
 {(f\psi_0)}^{-\frac{1}{p-1}}
W^{\frac{p}{p-1}} \lsim
 \vl. \begin{cases}
  {(1+s)}^{-\frac{2(p+1)}{p-1}} & \quad \text{for }
 s \geq 0 \\
  \e^{\frac{s}{2M}\frac{p-2}{p-1}} &
  \quad \text {for } s<0
 \end{cases} \vr\} \lsim {(1+|s|)}^{-2};
\eeq
 so, from the H\"older's inequality, we get
\begin{align*}
\int vW & \geq -\int |v|W \\
        & = -\int \( f^{\frac{1}{p}} |v| \psi_0^{\frac{1}{p}} \)
                  \( f^{-\frac{1}{p}} W \psi_0^{-\frac{1}{p}} \)
                  \\
        & \geq -U(t) {\( \int {(f\psi_0)}^{-\frac{1}{p-1}}
        W^{\frac{p}{p-1}} \)}^\frac{p-1}{p} \\
        & \gsim -U(t)
\end{align*}
and finally
$$
F''_0(t) \geq {U(t)}^p - CU(t)
$$
for a suitable positive constant $C$.

Now we are going to prove estimate \eqref{est-1}. Let observe that
 \beq \label{est.fpsi}
f^{-\frac{1}{p-1}} \psi_0 \lsim \vl\{ \begin{array}{ll}
{(1 + s)}^2 & \quad \text {for $s \geq 0$}\\
\e ^{\frac{s}{2M}\frac{p-2}{p-1}} & \quad \text {for $s<0$}
\end{array} \vr\} \lsim {(1 + |s|)}^2;
\eeq
proceeding as before, we obtain

\begin{gather*}
\begin{split}
F_1 (t) & \leq \e^{-t/2M} \int_{\ns R} |v| \phi_1 \di s \\
        & = \e^{-t/2M} \int_\dom \( f^{\frac{1}{p}} |v|
                    {\psi_0}^{\frac{1}{p}} \)
                 \( f^{-\frac{1}{p}}
                    {\psi_0}^{-\frac{1}{p}} {\phi_1}\) \di s\\
     & \leq U(t)
            {\( \e^{-\frac{t}{2M}\frac{p}{p-1}} \int_\dom f^{-\frac{1}{p-1}}
                {\psi_0}^{-\frac{1}{p-1}}
                {\phi_1}^{\frac{p}{p-1}} \di s \)}^{\frac{p-1}{p}},
\end{split}
\end{gather*}
where, setting $$\Psi={(f\psi_0)}^{-\frac{1}{p-1}}
{(\e^{-\frac{t}{2M}}\phi_1)}^{\frac{p}{p-1}},$$ we have
$$
 \Psi \lsim \vl\{
\begin{array}{ll}
 {(1+s)}^\frac{p-2}{p-1} \e^{\frac{s-t}{2M}\frac{p}{p-1}} &
    \quad \text{for $s \geq 0$}, \\
 \e^{\frac{s}{2M}\frac{p-2}{p-1}} \e^{-\frac{t}{2M}\frac{p}{p-1}} &
  \quad \text{for $s <0$}.
\end{array} \vr.
$$

Now we observe that for $s \rightarrow -\infty$, $\Psi$ decreases exponentially,
while as for as the behaviour at $\infty$ is concerned, we subdivide the domain of
integration into two disjoint parts:
\begin{align*}
D_0 & \doteq \{\dom\} \\
    & = \( D_0 \cap \{ s < t-t^{\epsilon'} \} \) \cup
      \( D_0 \cap \{ s \geq t-t^{\epsilon'} \} \) \\
    & = D_1^{\epsilon'} \cup D_2^{\epsilon'};
\end{align*}
we consider this partition for every $\epsilon'>0$ sufficiently small. If $s \in
D_1^{\epsilon'}$, we have $\e^{\frac{s-t}{2M}\frac{p}{p-1}} \leq
 \e^{-\frac{t^{\epsilon'}}{2M}\frac{p}{p-1}}$, thus $\Psi$ decreases esponentially
 as $s$ approaches to $\infty$. On the other hand, if $s \in D_2^{\epsilon'}$, we
 have only $s-t \leq s_0 + R$ and so $\Psi \lsim s^\frac{p-2}{p-1}$ is an
 increasing function on $D_2^{\epsilon'}$ as $s\rightarrow \infty$. So, modulo a
 positive constant, we can estimate the integral of $\Psi$ on $\{\dom\}$ with
 its integral on $D_2^{\epsilon'}$ (we consider $t$, and hence $s$, sufficiently large):
\begin{align*}
F_1(t) \lsim U(t) { \( \int_{D_2^{\epsilon'}} s^\frac{p-2}{p-1} \di s \)
}^\frac{p-1}{p}.
\end{align*}

But $D_2^{\epsilon'} \subset [t-t^{\epsilon'},t+s_0+R]$; moreover, for $\beta >
\alpha \geq 1$ and $\gamma \not= -1$, we have
$$
\int_\alpha^\beta s^\gamma \di s \lsim (\beta-\alpha)\alpha^\gamma,
$$
hence
$$
F_1(t) \lsim U(t) { \( \int_{t-t^{\epsilon'}}^{t+s_0+R} s^\frac{p-2}{p-1} \di s \)
}^\frac{p-1}{p} \lsim U(t){(t+R)}^{\epsilon'\frac{p-1}{p}} {(t+R)}^\frac{p-2}{p}.
$$

Remembering that $F_1(t) \gsim \epsilon$, setting $\epsilon_0=\epsilon'(p-1)$
and taking the power $p$ in both members, we get
\beq \label{est.Upa}
{U(t)}^p \gsim \epsilon^p {(t+R)}^{2-p-\epsilon_0};
\eeq
this is the first part of estimate \eqref{est-1}. Let note that argueing as in the proof of lemma \ref{l.Kn} to obtain \eqref{est.VV'T1}, we get $F_0 (t) \gsim \epsilon^p {(t+R)}^a \geq 0$ in $[0,T_1[$ and then in $[0,T[$.

To get the other part of \eqref{est-1}, we estimate $F_0$:
\begin{gather*}
\begin{split}
F_0 (t) & = \int_{\ns R} v\psi_0 \di s \\
        & = \int_{\dom} \( f^{\frac{1}{p}} v
                    {\psi_0}^{\frac{1}{p}} \)
                 \( f^{-\frac{1}{p}}
                    {\psi_0}^{\frac{p-1}{p}} \) \di s\\
     & \leq {\( \int_{\ns R} f |v|^p \psi_0 \di s \)}^{\frac{1}{p}}
            {\( \int_\dom f^{-\frac{1}{p-1}} \psi_0 \di s \)}
              ^{\frac{p-1}{p}} \\
     & \lsim U(t) {\( \int_\dom {(1+|s|)}^2 \di s \)}^{\frac{p-1}{p}}\\
     & \lsim  U(t) {(t+R)}^{\frac{3(p-1)}{p}};
\end{split}
\end{gather*}
taking the power $p$ in both sides gives
\beq \label{eq.Up1}
{U(t)}^p \gsim {(t+R)}^{-3(p-1)} {F_0 (t)}^p,
\eeq
and combining this inequality with \eqref{est.Upa}, we finally obtain needed estimate
 \eqref{est-1}.

We set $T_1(\epsilon) = \e^{C_0/\epsilon}$, $C_0>0$, and $s_0 (\epsilon) =
-(1+\delta_1) T_1(\epsilon)-R$, with $\delta_1 \in ]0,1[$ such
that $\delta_1 < M$; thanks to this choice, relation
\eqref{eq.lim} holds, hence it remains to prove estimate
\eqref{estF0''t1}  in $[0,T_1[$.

To begin, let observe that if $t \in [0,T_1[$, $\supp v \subset \{
s<0 \}$, since $s \leq s_0 + T_1 + R = -\delta_1 T_1$, thus we can
consider only the negative part in \eqref{est.fpsiW} and
\eqref{est.fpsi}.

Recalling what done to get \eqref{estF0''}, we obtain
\begin{align*}
\int vW & \geq -U(t) \( \int_\dom (f\psi_0)^{-\frac{1}{p-1}}
W^{\frac{p}{p-1}} \)^\frac{p-1}{p} \\
        & \gsim -U(t) \( \int \e^{\frac{s}{2M}\frac{p-2}{p-1}} \di
        s \)^{\frac{p-1}{p}} \\
        & \gsim -U(t) \e^{\frac{s_0+T_1+R}{2M}\frac{p-2}{p}} =
            -U(t) \e^{-\delta T_1};
\end{align*}
so, if we set
$$
\delta \doteq \frac{\delta_1 (p-2)}{2Mp} \in ]0,1[,
$$
from \eqref{F0''nonlin} we finally deduce
$$
F_0''(t) \gsim {U(t)}^p -\e^{-\delta T_1} U(t),
$$
which is estimate \eqref{estF0''t1}. So we can apply lemma
\ref{l.Kn} and get that $F_0$ blows up in finite time.

Now, considering the asymptotic behaviours of $\psi_0$, we obtain
$$
\psi_0(s) \lsim \vl\{ \begin{array}{ll}
1+s & \quad \text{if } s \geq 0 \\
\e^\frac{s}{2M} & \quad \text{if } s < 0\\ \end{array} \vr\} \lsim
1+|s|
$$
and hence, thanks to the Cauchy--Schwartz inequality, we deduce
$$
F_0 (t) \lsim \int_{\dom} (1+|s|)|v(t,s)| \di s \lsim {(t+R)}^{3/2} ||v(t)||_{\cs
L^2 (\ns R)}.
$$
Thus also $v$ blows up in finite time and this concludes the proof, assuming the statements of the following
section.

\section{Blow up with small initial data for $p \in ]1,2]$} 

First of all, we consider the case $p \in ]1,2[$; we want to apply the classical Kato's lemma \ref{l.K} with $q=3(p-1)$ and $a = 4-p-\epsilon_0$, where $\epsilon_0 > 0$ sufficiently small. In particular, we assume
\beq \label{epspicc}
\epsilon_0 < 2-p.
\eeq
In this way, the relation $(p-1)a > q-2$ is true also for $p \in ]1,2[$.

We also assume that $s_0>0$ is very far from the black hole (in particular
\eqref{eq.sc0} still holds) and we consider only the region $\ns R^+ \doteq \{s \in \ns R^3 \; ; \; s > 0\}$. In this region, estimates \eqref{estF0''} and \eqref{est.Upa} still hold and thanks to our assumptions, \eqref{est.Upa} implies that definitely $U(t) \geq 1$, i.e.
$$
F_0''(t) \gsim {U(t)}^p \gsim \epsilon^p{(t+R)}^{2-p-\epsilon_0}.
$$
Integrating twice, we get
$$
F_0(t) \gsim \epsilon^p {(t+R)}^{4-p-\epsilon}.
$$
But also \eqref{eq.Up1} still holds, hence we have
$$
F_0''(t) \gsim {(t+R)}^{-3(p-1)}{F_0(t)}^p,
$$
and so we can apply the lemma, provided the solution lives in $\ns R^+$. But this is true, since
$$
U(t) \gsim \epsilon {(t+R)}^\frac{2-p-\epsilon_0}{p} \geq 1
$$
if and only if
$$
t + R \geq { \( \frac{1}{\epsilon} \) }^\frac{p}{2-p-\epsilon_0}
$$
and the quantity in the right side is polinomial in $\epsilon$, while condition  \eqref{eq.sc0} guarantees that $s_0(\epsilon)$ increases more rapidly than every polinomial as $\epsilon \rightarrow 0$; moreover, also $T(\epsilon)$ is polinomial in $\epsilon$, hence
$$
s \geq s_0(\epsilon) - t - R \geq  s_0(\epsilon) - T(\epsilon) - R > 0.
$$

It remains to consider the case $p=2$. In this situation, we do not need any hypothesis on $s_0(\epsilon)$.

\section{Appendix} \label{secteclemmas} 

Our first step in this section is to consider the problem \beq
\label{prH0} \vl\{
\begin{array}{ll}
-\phi^{\prime\prime} (s) +H(s)\phi(s) = 0 & \quad s \in \ns R, \\
|\phi(s) -bs| \lesssim \log (2+ s) & \quad \text{for }   s \geq 0, \\
0 < \phi (s) \lsim 1 & \quad \text{for $s<0$},
\end{array} \vr.
\eeq where the potential $H(s)$ is assumed to satisfy
\begin{equation}\label{eq.asHs}
   0 < H(s) \lesssim (1+|s|)^{-a}
\end{equation}
for some $a \geq 3$. Our first result is the following.

\begin{lemma} \label{lemfi0Hs}
There exists a real number $b>0$ such that the problem
\eqref{prH0} has  a positive solution $\phi_0 \in \fs C^2 (\ns R)$
such that the limit
$$
D = \lim_{s \rightarrow -\infty} \varphi_0(s) $$ exists, $D >0$
and the following relation
\begin{equation}\label{eq.impr2-a}
     \varphi_0(s) - D \sim {|s|}^{2-a} \qquad \text{for } s
     \rightarrow -\infty
\end{equation}
holds.
\end{lemma}

\prf Consider the  Cauchy problem \beq \label{prCPph} \vl\{
\begin{array}{ll}
-y^{\prime\prime} (s) +H(s)y(s) = 0 & \quad s \in \ns R, \\
y(0)= 1, \quad y^\prime(0)= 0.
\end{array} \vr.
\eeq This Cauchy problem has a unique solution $y(s) \in \fs C^2
(\ns R).$ A qualitative analysis of the equation and the
assumption $H(s) > 0$ show that the solution satisfies
$$  y^\prime(s) > 0 \qquad {\rm for} \   s >0, $$
$$  y^\prime(s) < 0 \qquad {\rm for} \  s < 0, $$ and hence
$$y(s) \geq 1$$ for all real $s.$

 One can rewrite
problem \eqref{prCPph} in the form of the following integral
equation
\begin{equation}\label{eq.intphi}
    y(s) = 1 + I(y)(s),
\end{equation}
where
\begin{equation}\label{eq.dI}
    I(y)(s) = \int_{0}^s \int_{0}^{\sigma} H(\theta)
    y(\theta) \di \theta \di \sigma = \int_{0}^s (s-\theta) H(\theta)
    y(\theta) \di \theta.
\end{equation}
We shall  show that
\begin{alignat}{2}
    & |y(s) -d_+ s|\lesssim \log (2+s) & \qquad & \text{for } s \geq 0,
    \label{eq.bound+}\\
    & |y(s) -d_- s|\lesssim \log (2+|s|) & \qquad & \text{for } s < 0,
    \label{eq.bound-}
\end{alignat}
where
 $$
d_\pm = \int_{0}^{\pm \infty}  H(\theta)
    y(\theta) \di \theta.
 $$

The assumption \eqref{eq.asHs} shows that the integral operator
$I(y)(s)$ is well--defined and satisfies the estimate
\begin{equation}\label{eq.linf}
0 \leq  I(y)(s) \leq s \int_{0}^s H(\theta)
    y(\theta) \di \theta;
\end{equation}
hence \eqref{eq.intphi} implies the inequality
\begin{equation}\label{eq.linfjk}
y(s) \leq  1 + s \int_{0}^s H(\theta)
    y(\theta) \di \theta.
\end{equation}

Now, let consider the case $s \geq 0$. The  previous inequality
yields
\begin{equation}\label{eq.boundr}
    y(s) \leq 1 + s\int_0^\infty H(\theta) y(\theta) \di \theta = 1 + d_+ s.
\end{equation}

On the other hand, combining \eqref{eq.intphi} and \eqref{eq.boundr}, we get
\begin{align*}
y(s) - 1 - d_+ s & = s\int_0^\infty H(\theta) y(\theta) \di \theta
                       - s\int_s^\infty H(\theta) y(\theta) \di \theta \\
       & \qquad -\int_0^s \theta H(\theta) y(\theta) \di \theta  - d_+ s \\
       & = - s\int_s^\infty H(\theta) y(\theta) \di \theta
             -\int_0^s \theta H(\theta) y(\theta) \di \theta  \\
       & \gsim  -s\int_s^\infty \frac{d\theta}{{(1+\theta)}^{a-1}}
                       -\int_0^s \frac{d\theta}{{(1+\theta)}^{a-2}}.
\end{align*}
But, for each $a \geq 3$, we have also
\begin{gather*}
\int_s^\infty \frac{d\theta}{{(1+\theta)}^{a-1}} \lsim {(1+s)}^{2-a}, \\
\int_0^s \frac{d\theta}{{(1+\theta)}^{a-2}} \lsim \log (2+s),
\end{gather*}
and thus we deduce
 $$
 y(s) - 1 -  d_+ s \gsim - \log (2+s).
 $$
From this equation and \eqref{eq.boundr}, we finally obtain the precise asymptotic
 estimate \eqref{eq.bound+}. Analogously, we get the parallel result for
 $s<0$. It is important to note that $d_+ > 0$ and
 $$d_- = -\int_{-\infty}^{0}  H(\theta)
    y(\theta) \di \theta < 0.$$

In a similar way, we can consider the Cauchy problem \beq
\label{prCz} \vl\{
\begin{array}{ll}
-z^{\prime\prime} (s) +H(s)z(s) = 0 & \quad s \in \ns R, \\
z(0)= 0, \quad z^\prime(0)= 1.
\end{array} \vr.
\eeq
Obviously, this Cauchy problem has a unique solution $z(s) \in
\fs C^2 (\ns R).$ The assumption $H(s) > 0$ guarantees that the
solution satisfies
$$  z^\prime(s) > 0 \qquad \forall s  \in \ns R, $$
so
$$  z(s) > 0 \qquad {\rm for} \  s > 0 $$ and
$$  z(s) < 0 \qquad {\rm for} \  s < 0. $$

The integral equation \eqref{eq.intphi} has to be replaced by
\begin{equation}\label{eq.intz}
    z(s) = s + I(z)(s)
\end{equation}
and the argument given in the proof of estimates \eqref{eq.bound+}
and \eqref{eq.bound-} leads to
\begin{alignat}{2}
    & |z(s) -e_+ s| \lsim \log (2+s) & \qquad & \text{for } s \geq
    0, \label{eq.boundz+} \\
    & |z(s) -e_- s| \lsim \log (2+|s|) & \qquad & \text{for } s <
    0, \label{eq.boundz-} \\
\end{alignat}
where
$$
e_\pm = 1 + \int_{0}^{\pm \infty}  H(\theta) z(\theta) \di \theta.
$$
Note that $e_+ > 0$ and $$ e_- =  1 - \int_{-\infty}^{0}
H(\theta)
    z(\theta) \di \theta  > 0.$$

Setting
$$ \phi(s) = e_- y(s) - d_- z(s), \qquad b = e_-d_+ - d_- e_+ > 0,
$$
we take advantage of \eqref{eq.bound-} and \eqref{eq.boundz-}, and
conclude that
\begin{equation}\label{eq.limp}
   | \phi(s)| \lesssim \log (2+|s|)   \qquad \text{for }   s  < 0,
\end{equation}
while from \eqref{eq.bound+} and \eqref{eq.boundz+}, we deduce
\begin{equation}\label{eq.estphi-}
    |\phi(s) - b s| \lesssim \log(2+s)     \qquad   \text{for } s  >
    0.
\end{equation}

To improve estimate \eqref{eq.limp}, we note that $\varphi(s)$
satisfies the integral equation
\begin{equation}\label{eq.intzphi}
    \varphi(s) = \varphi(0) +  s\varphi'(0) + I(\varphi)(s).
\end{equation}
As before, we have (for any $s < 0$)
\begin{align}\nonumber
\varphi(s) - \varphi(0) - \varphi^\prime(0)  s & =
-s\int_{-\infty}^0 H(\theta) \varphi(\theta) \di \theta
                       + s\int_{-\infty}^s H(\theta) \varphi(\theta) \di \theta \\
       & \qquad +\int_{-\infty}^0\theta H(\theta) \varphi(\theta) \di \theta
-\int_{-\infty}^s\theta H(\theta) \varphi(\theta) \di \theta
\label{eq.simph}
\end{align}
and then a combination between \eqref{eq.limp} and the assumption
\eqref{eq.asHs} implies
\begin{equation}\label{est.1ost}
    s\int_{-\infty}^s H(\theta) |\varphi(\theta)| \di \theta \lsim
(1+|s|)^{2-a} \log(2+|s|) \lesssim 1,
\end{equation}
\begin{equation}\label{est.2ost}
   \int_{-\infty}^s\theta H(\theta) |\varphi(\theta)| \di \theta \lsim
(1+|s|)^{2-a} \log(2+|s|) \lesssim 1,
\end{equation}
so \eqref{eq.simph} yields
\begin{align}\nonumber
\left| \varphi(s) - \left(\varphi^\prime(0) -  \int_{-\infty}^0
H(\theta) \varphi(\theta) \di \theta \right)  s \right| \lesssim
1.
\end{align}
Comparing this estimate with \eqref{eq.limp}, we see that
$$
\varphi^\prime(0) -  \int_{-\infty}^0 H(\theta) \varphi(\theta)
\di \theta  = 0;$$ so, setting
\begin{equation}\label{eq.posD}
    D = \varphi(0) + \int_{-\infty}^0\theta H(\theta) \varphi(\theta) \di \theta ,
\end{equation}
 we use \eqref{eq.simph}, \eqref{est.1ost}, \eqref{est.2ost} and
 we arrive at
\begin{equation}\label{eq.limpimp}
   | \phi(s)| \lesssim 1  \qquad {\rm for } \   s  < 0
\end{equation}
and \eqref{eq.impr2-a}.

 The function $\phi(s)$ is obviously
positive near $s=0.$ Moreover, for $s>0$, $\phi(s)$ increases and
is positive. It is easy to show that $\phi(s)\geq 0$ for all
$s<0.$ Indeed, if $\phi(s_0)<0$ for some $s_0<0,$ then $\phi(s_1)
< 0$, $\phi^\prime(s_1) > 0$ for some $s_1 < 0$, thus the equation
$$ \phi^{\prime\prime}(s) = H(s) \phi(s)$$ implies that $$\phi(s)
< 0, \qquad \phi^\prime(s) > 0, \qquad \phi^{\prime\prime}(s) <0
\qquad {\rm for } \ s < s_1.$$ This contradicts \eqref{eq.limpimp}
and shows that $\phi(s)\geq 0$ for all $s<0.$

The positiveness of $D$ follows from the fact that $\varphi(0)
>0,$  $\varphi(s)
\geq 0$ for all $s \in \ns{R}$ and \eqref{eq.posD}.

To complete the proof, let observe that we have indeed $\phi \geq
D$ (thus $\phi \sim 1$ as $s \rightarrow -\infty$ and $\phi >0$ in
$\ns R$), since
$$
\phi(s) - D = \int_{-\infty}^s (s-\theta) H(\theta) \phi (\theta)
\di \theta \geq 0;
$$
from this relation and \eqref{eq.asHs}, we also get
 $$
 \phi(s)-D \sim \int_{-\infty}^s
 \frac{s-\theta}{{(1+|\theta|)}^{a}}\di \theta
 \sim {|s|}^{2-a} \qquad \text{for } s \rightarrow -\infty
 $$
and this prove \eqref{eq.impr2-a}.\bs

\begin{lemma} \label{lemfi0}
There exists a positive function $\phi_0 \in \fs C^2(\ns R)$ such
that $G \phi_0 = 0$ in $\ns R$ and for some positive constants $b$
and $D$ we have
$$
\vl\{ \begin{array}{ll}
|\phi_0 (s) -bs| \lsim \log (2+|s|) & \quad \text{for } s \geq 0, \\
\phi_0 (s)-D \sim \e^{s/2M} & \quad \text{for } s \rightarrow
-\infty.
\end{array} \vr.
$$
\end{lemma}

\prf Let $\phi$ satisfy
$$
\phi'' (s) -\frac{2MF}{{r(s)}^3} \phi (s)=0.
$$
Using the asymptotic estimates \eqref{eq.estrs}, \eqref{eq.estFs}
of lemma \ref{lm,asrfl}, we find

\beq \label{relFr}
\frac{F}{r^3}
\sim \begin{cases}
s^{-3} & \quad \text{if } s \rightarrow \infty, \\
\e^{s/(2M)} & \quad \text{if } s \rightarrow -\infty
\end{cases}
\eeq and the claim follows  from the argument of the proof of the
previous lemma (with $a=3$). The main modification concerns
relation \eqref{est.1ost}. In fact the asymptotic expansion
\eqref{relFr} for $s < 0$ implies that
\begin{align} \begin{split} \label{est.1ostm}
    \phi(s) - D & =
    \int_{-\infty}^s (s-\theta) H(\theta) \varphi(\theta) \di
    \theta \\
    & \sim  \int_{-\infty}^s (s-\theta) \e^{\theta/2M} \di \theta
    = \frac{\e^{s/(2M)}}{4M^2} \end{split}
\end{align}
as $s \rightarrow -\infty$, and this leads to
 $$
\phi_0 (s)-D \sim \e^{s/2M} \qquad \text{for } s \rightarrow
-\infty.
 $$
The rest of the claim follows directly from the assertion of the
previous lemma.\bs

Now we state a corollary of the Levinson's theorem (see
 \cite{Fed}, page 49, chapter 2 \S 5.4), which we
shall apply to get the estimate for $\phi_1$.

\begin{prop} \label{proplev}
Consider the equation \beq \label{eqlev} y^{(n)} + \sum_{k=1}^n
\alpha_k (s) y^{(n-k)} = 0, \qquad s \in \ns R^+, \eeq where
$\alpha_k (s) \in \fs C^{\infty} (\ns R^+)$ are complex--valued
functions such that
$$
\alpha_k (s) = \beta_k + \gamma_k (s),
\qquad \int_{\ns R^+} |\gamma_k (s)| \di s < \infty,
$$
and let $q_1, q_2, \ldots, q_n$ be the distinct roots of the equation
$$
q^n + \sum_{k=1}^n \beta_k q^{n-k} = 0.
$$

Then equation (\ref{eqlev}) has $n$ linearly independent solutions
$$ y_j (s) , \qquad j=1,2,\ldots, n, $$
having the asymptotic expansion
$$
y_j^{(k-1)} (s) = q_j^{k-1} \e ^{q_j s} [1 + o(1)]
\qquad \text{as } s \rightarrow \infty,
$$
where $j,k=1,2,\ldots, n$.
\end{prop}

\begin{lemma} \label{lemfi1}
Given any $A>0,$ the equation \beq \label{probfi1} (G+A^2)\phi(s) = 0, \qquad s \in
\ns R \eeq admits a positive solution $\phi_1 \in \fs C^2 (\ns R)$ such that $\phi_1
(s) \sim \e ^{As}$ as $|s|$ approaches $\infty$.
\end{lemma}

\prf According to proposition \ref{proplev} and relation
\eqref{relFr}, there exists a solution $\phi_1$ of \eqref{probfi1}
such that $\phi_1(s) \sim \e^{As}$ as $s \rightarrow -\infty$.
From
$$
\phi_1'' = \( \frac{2MF}{r^3} +A^2 \) \phi_1, \qquad
\frac{2MF}{r^3} +A^2 >0
$$
and a qualitative study, we get \beq \label{estfi1} \phi_1''(s) >
0 \qquad \text{and thus } \phi_1(s) > 0 , \ \phi_1^\prime(s) >
0\eeq for each $s \in \ns R$. Now, from  Proposition \ref{proplev}
for $s \rightarrow +\infty$,
  we deduce $\phi_1 (s) \sim \lambda \e^As + \mu \e^{-As}$ for
 suitable $\lambda$, $\mu \in \ns R$ and $s \rightarrow +\infty$.
 The property \eqref{estfi1} guarantees that
 $\lambda>0,$ so
  it is necessarily $\phi_1 (s) \sim \e^{As}$ for $s \rightarrow +\infty$ and
  the proof is finished.
\bs

\begin{lemma} \label{lemF1}
Let $F_1$ be defined as in (\ref{defF1}). Then $F_1 (t) \gsim
\varepsilon$ holds for all $t \geq 0$.
\end{lemma}

\prf We multiply the  equation (\ref{probv})  by $\psi(t,s) \doteq
\e^{-t/2M} \phi_1 (s)$ and integrate over $\ns R$ in $s$ and over
$[0,\tau]$ in $t$:
$$
\int_0^\tau \int_{\ns R} \( v_{tt} - v_{ss} +\frac{2MF}{{r(s)}^3} v\)
            \psi \di s \di t
    = \int_0^\tau \int_{\ns R} f |v|^p \psi \di s \di t.
$$
Note that the initial data in \eqref{probv} are compactly
supported due to \eqref{ipdi} and a finite dependence domain
argument for the equation \eqref{probv} implies that $v(t,s)$ has
compact support in $s$ for $t$ bounded. Further, we have the
following regularity assumption $$v \in \cap_{k=0}^2  \fs C^k
([0,T[;\cs H^{2-k} (\ns R))$$ for the solution of \eqref{probv}, so
we can apply an integration by parts argument and obtain
\begin{align*}
& - \int_0^\tau \int_{\ns R} v\( \psi_{tt} - \psi_{ss} +\frac{2MF}{{r(s)}^3}
  \psi \) \di s \di t
         + \int_0^\tau \int_{\ns R} f |v|^p \psi \di s \di t \\
& \qquad = \int_{\ns R} (v_t\psi - v\psi_t) \di s \pp|_{t=\tau}
  - \int_{\ns R} (v_t\psi - v\psi_t) \di s \pp|_{t=0} .\end{align*}
  The right side of this equality can be rewritten as
  \begin{align*}
 & \int_{\ns R} (v_t\psi - v\psi_t) \di s \pp|_{t=\tau}
  - \int_{\ns R} (v_t\psi - v\psi_t) \di s \pp|_{t=0} \\
& \qquad = \int_{\ns R} (v_t\psi + v\psi_t) \di s \pp|_{t=\tau}
  - 2\int_{\ns R} v \psi_t \di s \pp|_{t=\tau}
  - \int_{\ns R} \e^{-t} (v_t\phi_1 + v\phi_1) \di s \pp|_{t=0} \\
& \qquad = \frac{d}{d\tau} \int_{\ns R} v\psi \di s
  + 2\int_{\ns R} v\psi \di s
  - \int_{\ns R} (v_0 + v_1) \phi_1 \di s
\end{align*}
due to the property $ \psi_t = -\psi/2M.$  The relation
$$\psi_{tt} - \psi_{ss} + 2MF{r(s)}^{-3} \psi = \e^{-t/2M}
\(G+\frac{1}{4M^2}\)\phi_1  = 0$$ implies
$$- \int_0^\tau \int_{\ns R} v\( \psi_{tt} - \psi_{ss}
+\frac{2MF}{{r(s)}^3}
  \psi \) \di s \di t =0;$$
  so, being
  $$
F_1 (t) = \e^{-t/2M} \int_{\ns R} v(t,s) \phi_1 (s) \di s,
$$
we arrive at
$$
F_1'(\tau) + 2F_1(\tau) = \int_0^\tau \int_{\ns R} f |v|^p \psi \di s \di t +
\int_{\ns R} (v_0 + v_1) \phi_1 \di s. $$

The right side of this identity is greater than $\epsilon$
multiplied by a positive constant (since $\phi_1>0$ and $\int v_j
\di s \geq \epsilon$, $j=0,1$), so we get
$$
F_1'(\tau) + 2F_1(\tau) \gsim \varepsilon.
$$
Now, we multiply both sides by $\e^{2\tau}$ obtaining
$$
\frac{d}{d\tau} \( \e^{2\tau} F_1(\tau) \) \gsim \e^{2\tau}
\varepsilon
$$
and integrating on $\tau \in [0,t]$ we deduce
$$
\e^{2t}F_1(t) \gsim F_1(0) + \varepsilon(\e^{2t} -1),
$$
namely
$$
F_1(t) \gsim F_1(0) \e^{-2t} + \varepsilon - \varepsilon \e^{-2t}
\gsim \varepsilon,
$$
because $F_1(0)=\int v_0 \phi_1 >0$.\bs

\ \ \bibliographystyle{amsalpha} 

\end{document}